\documentclass[preprint,12pt]{iopart}
\usepackage[pdftex]{graphicx}
\usepackage{hyperref}
\usepackage{textcomp}
\pdfoutput=1
\hypersetup{pdfauthor={some author},pdftitle={eye-catching title}}
\ifpdf

\begin{document}

\title{Instability of some divalent rare earth ions and photochromic effect}
\author{A.V. Egranov$^{1,2}$, T.Yu. Sizova$^1$,   R.Yu. Shendrik$^{1,2}$ and N.A. Smirnova$^1$}

\address{$^1$ Vinogradov Institute of Geochemistry, Russian Academy of Sciences, Favorskii street 1a, 664033 Irkutsk, Russia}
\address{$^2$ Irkutsk State University, Faculty of Physics,  Gagarin Blvd. 20, 664003 Irkutsk, Russia}
\ead{alegra@igc.irk.ru}

\begin{abstract}
It was shown that the divalent rare earth ions (La, Ce, Gd, Tb, Lu, and Y) in cubic sites in alkaline earth fluorides are unstable with respect to electron autodetachment since its d$^1$(e$_g$) ground state is located in the conduction band which is consistent with the general tendency of these ions in various compounds. The localization of doubly degenerate d$^1$(e$_g$) level in the conduction band creates a configuration instability around the divalent rare earth ion that leading to the formation of anion vacancy in the nearest neighborhood, as was reported in the previous paper [Journal of Physics and Chemistry of Solids 74 (2013) 530-534]. Thus, the formation of the stable divalent ions as La, Ce, Gd, Tb, Lu, and Y (PC$^+$ centers) in CaF$_2$ and SrF$_2$ crystals during x-ray irradiation occurs via the formation of charged anion vacancies near divalent ions (Re$^{2+}$v$_a$), which lower the ground state of the divalent ion relative to the conductivity band. Photochromic effect occurs under thermally or optically stimulated electron transition from the divalent rare earth ion to the neighboring anion vacancy and reverse under  ultraviolet light irradiation.
\end{abstract}

\pacs{61.72.Cc,61.80.-x,71.55.Ht,71.70.Ej,78.55.Hx}% PACS, the Physics and Astronomy % Classification Scheme.

\maketitle

\section{Introduction}
The rare earth elements (Re) include the elements scandium, yttrium, and the lanthanides (Ln) (lanthanum through lutetium). Most of the rare earths in the solid state are trivalent; this is in contrast to the atomic state, where the rare earths are divalent. However, some compounds involving Eu, Sm, Tm and Yb can become divalent. Ce compounds, on the other hand, can be either trivalent or tetravalent. The synthesis of the non-classical divalent complexes was once deemed impossible \cite{Meyer:2012a,Cotton:2006,Meyer:1992}.

Dihalides of the rare earth elements became known shortly after the turn of the twentieth century. A first picture was completed in the late 1920s when it was thought that only the classical four, europium, ytterbium, samarium, and thulium, could be obtained in the divalent state. 

Lanthanide compounds with the lanthanide in the +2 oxidation are known for chalcogenides and halides. Binary examples are EuO and TmS as well as LaI$_2$ and NdCl$_2$. The compounds based on  La, Ce, Gd, Pr, Gd ions do not generate a stable divalent state with oxygen, fluorine, chlorine and bromine  ions. The ability to form a compound containing divalent rare earth ions are best analyzed by the formation iodide as iodide ion (I$^-$) is  the best reductant among F$^-$, Cl$^-$, Br$^-$ and I$^-$. Divalent iodides Nd, Sm, Eu, Dy, Tm, Yb are good electrical insulators, and form stable compounds whereas LaI$_2$, CeI$_2$, PrI$_2$, GdI$_2$ have a metallic luster and  good electronic conductivity \cite{Cotton:2006}. 

 Two cases of electronic configurations have to be distinguished, [Xe]6s$^0$5d$^0$4f$^n$ and [Xe]6s$^0$5d$^1$4f$^{n-1}$.The former with electrons occupying only 4f orbitals are salt-like insulators and their crystal chemical behavior is very similar to that of the respective alkaline-earth compounds. In the latter, there is one voluminous, outer-shell 5d orbital occupied. This 5d electron might be localized and involved in chemical bonding in cluster complexes. It may also be delocalized into a 5d band causing semiconducting or metallic behavior \cite{Meyer:2012a}.

Thus, by the beginning of the 21st century, the impression is that only traditional rare earth ions, whose number increased to six (Nd, Sm, Eu, Dy, Tm, Yb) may form a divalent state (for example \cite{Cotton:2006}).

It was thus a great surprise when it was reported in 2008 that the 5d$^1$ configuration could be preserved in an organometallic coordination compound \cite{Hitchcock:2008,Meyer:2008}. It recently has been shown that Re$^{2+}$ ions are accessible for all of the lanthanides except Pm, which was not investigated because of its radioactivity. Complexes of nine new +2 ions, La$^{2+}$, Ce$^{2+}$, Pr$^{2+}$, Gd$^{2+}$, Tb$^{2+}$, Ho$^{2+}$, Er$^{2+}$, Y$^{2+}$, and Lu$^{2+}$, were synthesized (\cite{Hitchcock:2008,Meyer:2008,Meyer:2014,Fieser:2015} and references there in). The synthesis of these compounds opens a new area of organometallic chemistry of lanthanides as d elements. The study of d$^1$ lanthanide inorganic compounds has already produced exciting results for materials science \cite{Simon:2006} and now can be extended to molecular organolanthanide complexes \cite{Hitchcock:2008}. 

However, it was not a surprise when McClure  and Kiss \cite{McClure:1963,Wybourne:1965} have succeeded in reducing all the lathanides (except Pm and Lu) to the divalent state by $\gamma$-irradiation of CaF$_2$ doped with trivalent rare earth ions. Later it was shown that not all bands can be attributed to the absorption of the divalent rare earth ions \cite{Nugent:1973,Merz:1967}. Further the so called photochromic centers were obtained either by x - ray irradiation or by additive coloration (heating the crystals in a calcium atmosphere) of CaF$_2$ crystals doped with certain rare earths ions (La, Ce, Gd, Tb, Lu or Y) which have low third ionization potentials. On the basis of optical and electron paramagnetic resonance ({EPR}) work, as well as theoretical investigations, it has been suggested that the ionized and thermally stable photochromic centers in {CaF$_2$} crystals consist of one and two electrons bound at an anion vacancy adjacent to a trivalent impurity cation and they were called as PC$^+$ and PC respectively \cite{Staebler:1971,Alig:1971,Anderson:1971}. It can be assumed that the formation of the photochromic centers in alkaline earth fluorides consistent with the general trend of these rare earth ions do not create stable compounds in the divalent state. Nevertheless, the possibility of the production of these ions in the divalent state in the cubic environment is persisted for several reasons \cite{Alig:1969,Buskes:1974}. Therefore, there was a contradictory situation. On the one hand, the photochromic centers are formed in {CaF$_2$} doped with the rare earth ions, which are unstable in the divalent state. On the other hand, there are a number of experimental results showing the formation of these ions in the divalent state in {CaF$_2$} (the best example is the formation of Ce$^{2+}$ in {CaF$_2$} \cite{Alig:1969}).

The purpose of this paper is to resolve this contradiction. It will show that the PC$^+$ center is a more stable configuration for the divalent rare earth ion than a cubic environment. The structure of the center can be represented as the divalent ion near an anion vacancy (Re$^{2+}$v$_a$, where Re - rare earth ion). Finally, it is discussed the role of the ligands in the formation of the stable divalent state for these ions. In previous work \cite{Egranov:2013} we have examined the mechanism of formation of these centers, which differs significantly from the other processes of the formation of impurity centers having in its composition of anion vacancies  \cite{Egranov:2008,Egranov:2008a,Egranov:2014}. In this article, we  also consider the mechanism of formation of the photochromic centers and clarify their structural model. Preliminary results have been published in the conference proceedings \cite{Egranov:2015}.

\section{Experimental}
Crystals of alkaline earth fluorides doped with rare earth ions (La, Ce, Gd, Tb, Lu, and Y) were synthesized using the Stockbarger technique from a melt in an inert atmosphere. Cadmium fluoride was added to the charge in order to prevent the formation of oxygen
impurities. Radiation coloration was performed using an X-ray tube with a Pd anode at 20 mA, 40 kV, and exposure times of $<$60 min. The absorption spectra were obtained on the Lambda 950 UV/VIS/NIR spectrophotometer at the Baikal Analytical Center for Collective Use, Siberian Branch, Russian Academy of Sciences.

\section{Electron transitions and model of the centers}
The effect of different conditions on the stability of impurity ion valences and the changes of its valence is of both applied and fundamental importance. Trivalent ions (Y, La, Ce, Gd, Tb, and Lu) are not reduced to the divalent state; instead, they form photochromic centers. According to the current model, each center has a trivalent ion, an anionic vacancy, and one (PC$^+$ center) or two (PC center) electrons \cite{Anderson:1971,Staebler:1971,Alig:1971}.

\begin{figure}
\centering
\includegraphics[width=5.2in, height=3.6in]{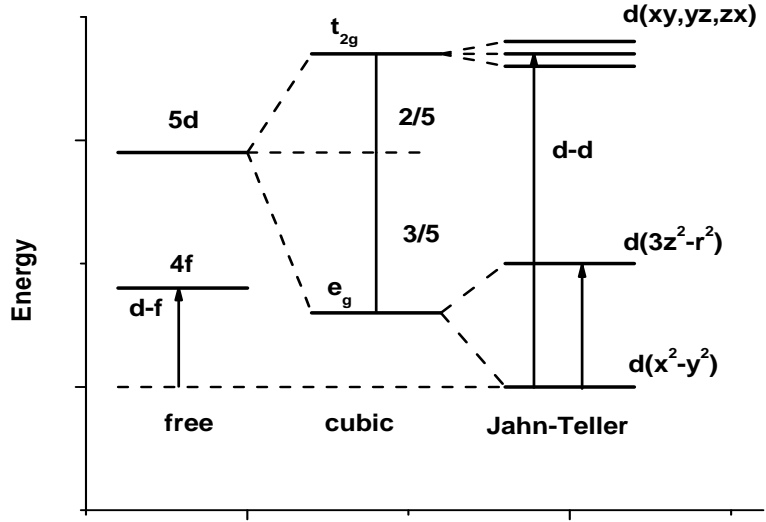}
\caption{Splitting of the levels of divalent ions in a crystal field.}
\label{state1}
\end{figure}

In a free state, La$^{2+}$, Gd$^{2+}$ Y$^{2+}$ ions have the ground state configuration d$^1$, free ions Ce$^{2+}$, Tb$^{2+}$ have f$^n$, but f$^{n-1}$5d$^1$ and f$^n$ are close in energy. For Lu$^{2+}$ the ground state configuration 4f$^{14}$6s is close in energy to 4f$^{14}$5d state. In alkaline earth fluoride crystals the cubic  crystal field splits the d state into two levels - doubly degenerate d(e$_g$) whose energy is reduced by 3/5$\Delta$ (where $\Delta$ - value of splitting of the d states by the crystal field) and the triply degenerate d(t$_{2g}$) state whose energy increases by 2/5$\Delta$, while the lower level is doubly degenerate \cite{Encyclopedia:2005} (Fig. ~\ref{state1}). Therefore, in many materials divalent ions Ce$^{2+}$, Tb$^{2+}$ and Lu$^{2+}$ have 4f$^1$5d$^1$, 4f$^8$5d$^1$ 4f$^{14}$5d$^1$ the ground state configuration, respectively. Thus, the ground state configuration of all these divalent ions is 5d$^1$ (or 4d$^1$ for Y$^{2+}$) and therefore in the absorption spectra two types of transitions can be expected - d${\rightarrow}$d transitions that should be similar for all of these ions and 5d$^1$ ${\rightarrow}$ f transitions for  La, Ce, Gd, Tb divalent ions from which is easy to define the valence state of the impurity (especially for Ce$^{2+}$  \cite{Alig:1969}). 

There are only  the stable photochromic centers (PC) after additive coloration of CaF$_2$ doped with trivalent rare earth ions (La, Ce, Gd, Tb, Lu, and Y) , which have been the object of careful study.  Ionized PC$^+$ centers were obtained from PC centers by ultraviolet bleaching. However, only part of the PC center was transformed to the PC$^+$ centers. This creates difficulties in studying of these centers so they are much less explored than PC centers. Recently, it has been shown that exposure by ionizing radiation of these crystals at a low temperature produces only PC$^+$ centers \cite{Bugaenko:2008,Sizova:2012} which upon heating (or optically) converted to PC centers.  Trivalent and divalent ions of yttrium have the simplest electronic structure with the empty d-shell and with 4d$^1$ electron configuration, respectively. The absorption spectrum of CaF$_2$-Y crystal x-ray irradiated at 80 K is shown in figure \ref{state2}.  It should be noted that the pure calcium fluoride crystal is radiation resistant  \cite{Hayes:1974a}. X- ray irradiation of CaF$_2$-Y crystals at 80 K results in the creation of PC$^+$ centers \cite{Bugaenko:2008} and the self-trapped holes (V$_k$ centers) with absorption band at about 4.0 eV \cite{Beaumont:1970}. Efficient creation of self-trapped holes indicates that the electron is captured by the trivalent rare earth ions and an unstable configuration of the divalent rare earth ion is created. Further the unstable configuration transforms into PC$^+$ centers by mechanism proposed by us in the previous paper \cite{Egranov:2013}. Since the initial state of the PC$^+$ has been the unstable divalent rare earth ion, we want to describe the transitions of  PC$^+$ center as d${\rightarrow}$d  and d${\rightarrow}$f transitions of the divalent rare earth ion.

\subsection{d ${\rightarrow}$ d Transitions.}
Substitutional divalent cations in CaF$_2$ occur at the centre of a cube of eight fluorines with cubic symmetry O$_h$. A d  level in this environment splits into an orbital triplet t$_{2g}$(xy,  yz,  zx) and  an orbital doublet e$_g$(x$^2$-y$^2$, 3z$^2$-r$^2$)  with an  energy separation $\Delta$=10Dq, and  the ground  state is e$_g$ in the case of the above rare earth ions.  The e$_g$ orbital degeneracy is not lifted by the spin-orbit coupling and it is expected a Jahn-Teller distortion of the ligand configuration will lift the degeneracy and  lower the energy (Fig. ~\ref{state1}). A d${\rightarrow}$d transition is forbidden when the surroundings are symmetric. Jahn–Teller lattice distortion partially allows optical transitions within the  d shell.

The splitting $\Delta$ of the d state of lanthanide in alkaline earth fluorides is of the order of 20000 cm$^{-1}$ \cite{Johnson:1969}  and  review of crystal field splitting of 5d-levels of trivalent Ce$^{3+}$ and divalent Eu$^{2+}$  in inorganic compounds has been made by Dorenbos \cite{Dorenbos:2002}.

The absorption band at ~2.3 eV (18550 cm$^{–1}$ ) (Fig. ~\ref{state2}) of CaF$_2$-Y most likely results of transitions from the lower e$_g$ to the t$_{2g}$ state, since the energy of the absorption peak is close to the d - splitting in the crystal field  (Fig. ~\ref{state1}). The absorption band has the unresolved structure, indicating poor splitting of the t$_{2g}$ state. The less intense absorption band at about ~1.5 eV (Fig. ~\ref{state2}) is most likely explained by optical transitions between the levels within the e$_g$ state splitting, due to the Jahn–Teller effect (Fig. ~\ref{state1}). 

In the first studies of CaF$_2$-Y in the 1960s \cite{O'Connor:1964}, the absorption bands observed after radiative coloration at 77 K and 300 K were attributed to absorption of Y$^{2+}$ ions with similar (but not the same) explanations of the optical transitions. It would seem, however, that this was not entirely correct, since it was later shown that these ions were not quite like other divalent rare earth ions. On the one hand, in contrast to other rare earth ions, they had no cubic symmetry; instead, they had  C$_{3v}$ symmetry. On the other hand, they had a clear photochromic effect, due to the presence of anionic vacancies \cite{Anderson:1971,Staebler:1971,Alig:1971}. The absorption bands were ascribed to absorptions of the PC and the PC$^+$ centers. It should be noted that for the both centers EPR signal was observed, attributed by O'Connor et al. \cite{O'Connor:1964} to divalent ions yttrium, which contradicts with the model of PC center having two paired electrons.

Since the d-state is the ground state for all the above ions, the same transitions should be expected in CaF$_2$ with other activators (La, Ce, Gd, Tb and Lu). On the figure ~\ref{state2} is shown the absorption spectrum of CaF$_2$-Gd x-ray  irradiated at 80 K which is very similar to that of CaF$_2$-Y as it would be expected from the d${\rightarrow}$d  transitions. Photoconductivity is observed in CaF$_2$-Gd crystals when excited by light with an energy of 2.0 eV \cite{Heyman:1969}; i.e., the t$_{2g}$ state occurs within the conductivity band.  

In some cases, the absorption spectrum is due to not only d${\rightarrow}$d transitions but as well as d${\rightarrow}$f transitions and the spectrum becomes more complicated. However, in these cases it is also possible to observe the absorption band 2.3 eV  associated with the transition within d shell (Fig. ~\ref{state3}a). The absorption spectra of crystals CaF$_2$-La and SrF$_2$-La x-ray irradiated at 80 K are similar to each other  and in their it can also be observed the bands associated with d${\rightarrow}$d transitions (Fig. ~\ref{state4}). Surprise is the shift of the absorption bands of SrF$_2$-La toward higher energies compared with that of CaF$_2$-La.

\subsection{d ${\rightarrow}$ f Transitions.}
\begin{figure}
\centering
\includegraphics[width=6.5in, height=4.5in]{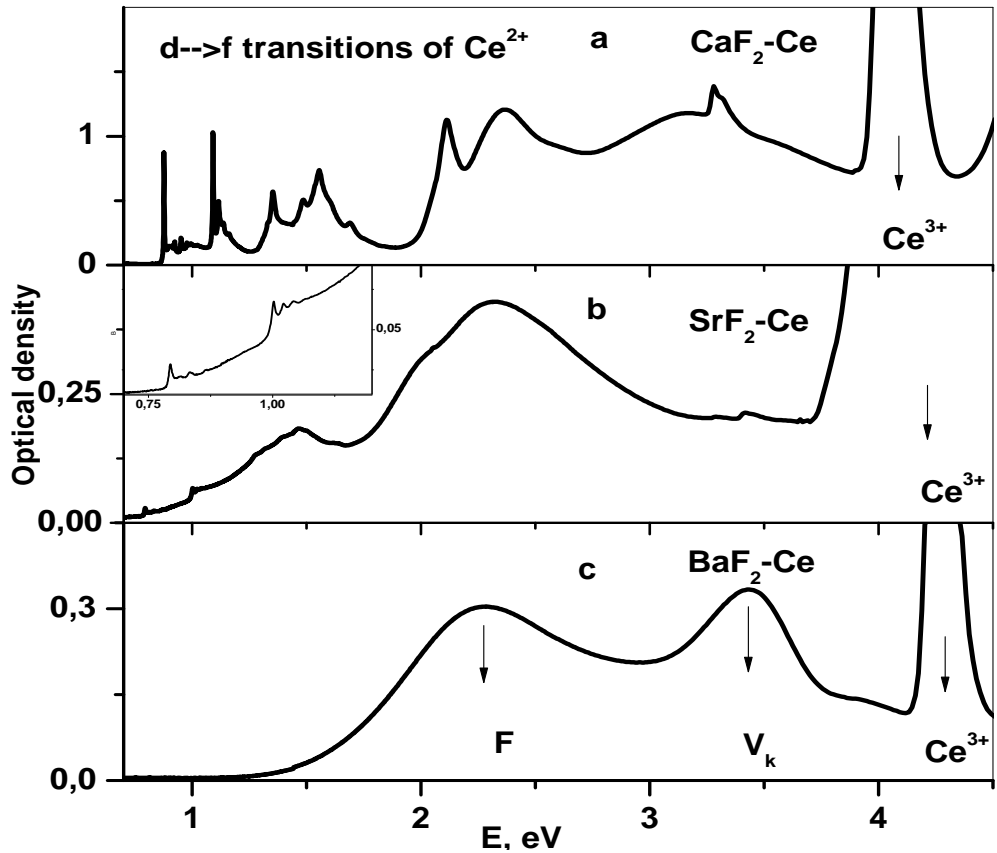}
\caption{Absorption spectra of  CaF$_2$–Ce (a), SrF$_2$-Ce (b) and BaF$_2$–Ce(c) crystals at 80 K after x-ray irradiated at 300 K (a) and 80 K (b),(c). The inset shows d ${\rightarrow}$ f   transitions of Ce$^{2+}$ in SrF$_2$}
\label{state3}
\end{figure}

In some cases the formation of PC$^+$ centers, either through the optical destruction of PC centers by ultraviolet light in additively colored calcium fluoride crystals \cite{Anderson:1971,Staebler:1971,Alig:1971} or through X-ray coloration of CaF$_2$ and SrF$_2$ crystals at 80 K \cite{Sizova:2012,Bugaenko:2008} was nevertheless accompanied by the formation of the divalent ions, as it was registered by optical \cite{Alig:1969} or  EPR measurements \cite{Hayes:1963,Herrington:1970}. An especially vivid absorption spectra were observed for divalent cerium ions \cite{Alig:1969}.
Fig. \ref{state3}a shows the absorption spectra of CaF$_2$-Ce crystals x-ray irradiated at room temperature ( in this crystal PC$^+$ center is stable at room temperature). Narrow lines associated with the  transitions of Ce$^{2+}$ are observed in the infrared. This line structure is also typical for SrF$_2$-Ce crystals x-ray irradiated at 80 K as shown in figure \ref{state3}b (in this crystal PC$^+$ centers is stable at low temperature only) \cite{Shendrik:2014}. The Ce$^{2+}$ centers in these crystals are destroyed under heating to room temperature.
The lowest d${\rightarrow}$f transition occurs at 7070 cm$^{-1}$ (0.88 eV)  and 6400 cm$^{-1}$ (0.79 eV) for CaF$_2$-Ce and SrF$_2$-Ce, respectively.

\begin{figure}
\centering
\includegraphics[width=5.2in, height=3.6in]{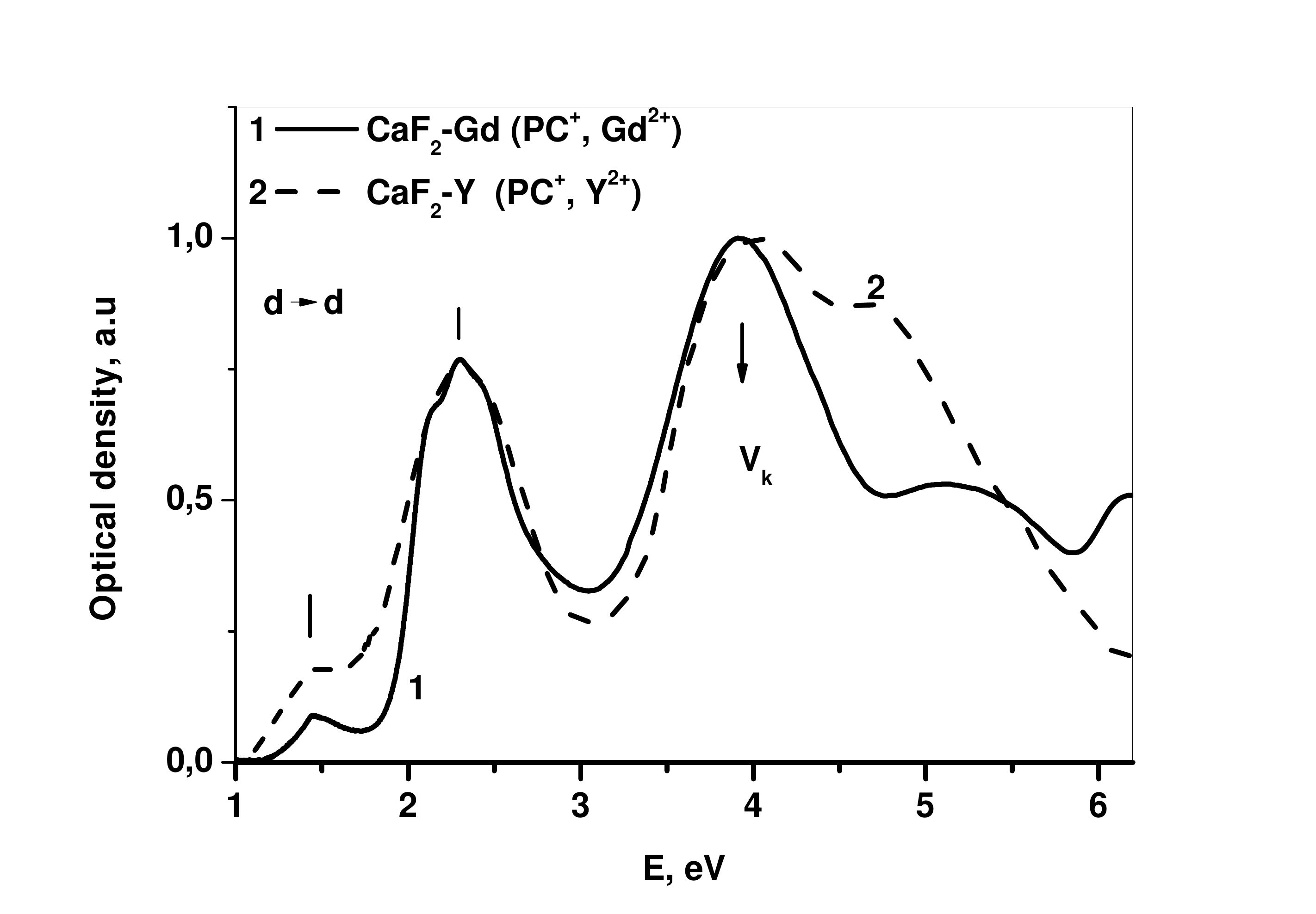}
\caption{{CaF$_2$-Gd} (1) and {CaF$_2$-Y} (2) crystals x-ray irradiated at 80 K.}
\label{state2}
\end{figure}

In CaF$_2$-La and SrF$_2$-La (Fig.~\ref{state4}) additional absorption bands (compared with CaF$_2$-Y Fig.~\ref{state2})  at 1.87 and 2.16 eV, respectively, can be attributed to d${\rightarrow}$f  transitions of La$^{2+}$.

PC$^+$ and divalent ions for the above rare earth ions in alkaline earth fluorides are always formed simultaneously, even though (according to the existing model) they are independent of one another. Neither PC$^+$ nor the divalent ions of these rare earth ions are formed in barium fluoride, although other rare earth divalent ions are formed during radiative coloration. Fig. ~\ref{state3}c shows the absorption spectra of BaF$_2$-Ce crystals x-ray irradiated at 80 K. It is evident from Fig. ~\ref{state3} that x-ray coloration results in the formation of F- and  V$_k$- centers, as in undoped barium fluoride crystals \cite{Nepomnyashchikh:2002}. However, the spectrum does not only show the absorption of PC$^+$ centers, but also there are no lines related to the divalent cerium ions in the infrared. All this indicates that the absorption bands of the divalent ions and PC$^+$ centers belong to the same defect.

At present there is a paradoxical situation. On the one hand, radiation or additive coloration of CaF$_2$ and SrF$_2$ crystals doped with trivalent ions (Y, La, Ce, Gd, Tb, and Lu) produces photochromic PC and  PC$^+$ centers, because the divalent states of these ions are unstable. On the other hand, EPR studies and the observed in the absorption spectra d ${\rightarrow}$ f   transitions clearly show that the divalent ions are also formed simultaneously with the PC$^+$ centers.

\subsection{Models for PC and PC$^+$ centers.} 
\begin{figure}
\centering
\includegraphics[width=6.5in, height=4in]{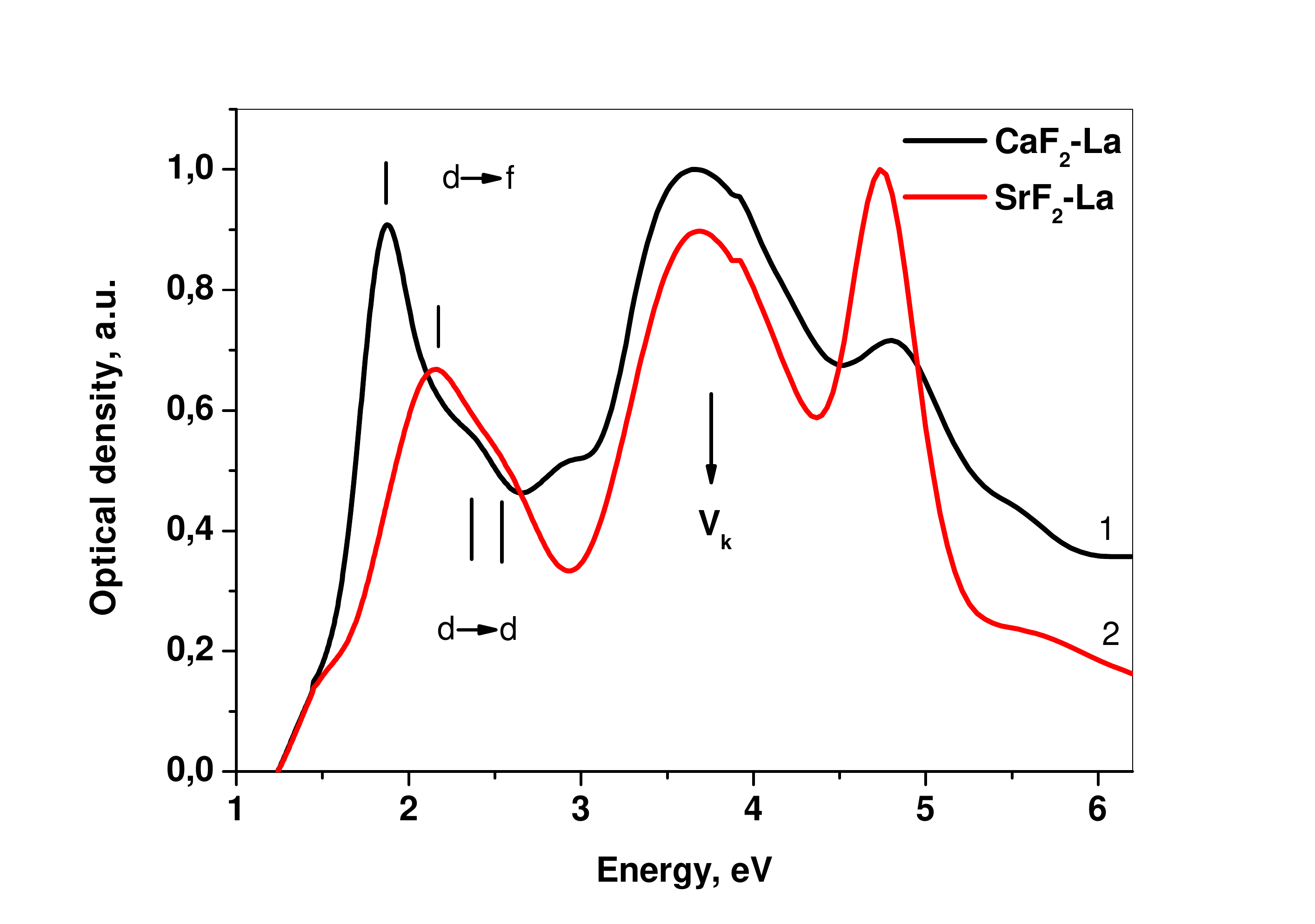}
\caption{Absorption spectra of  CaF$_2$–La (1), SrF$_2$-La (2) crystals at 80 K after x-ray irradiated at 80 K}
\label{state4}
\end{figure}

Thus, it can be assumed that the divalent rare earth ion is part of the PC$^+$ center, and all absorption bands can be explained by the d${\rightarrow}$d and d${\rightarrow}$f transitions in the divalent ions. On the other hand, PC$^+$ center contains anion vacancy, and the center structure can be envisioned as the divalent rare earth ion located beside an anion vacancy: Re$^{2+}$v$_a$ (Fig. ~\ref{state4}). The formal structure of the PC+ center is changed a little. Significant changes are related only to the electron localization. The photochromic effect results from the thermally or optically induced by red or infrared (IR) light transfer of electron from the divalent ions to the anion vacancy and vice versa by ultraviolet (UV) light. All of the experimental data can thus be explained by photo or thermal transformations within a single center Re$^{2+}$v$_a$, instead of three formally independent centers (PC,  PC$^+$, and divalent ions  Re$^{2+}$). 

It can be assumed that the metastable state Re$^{2+}$v$_a$ is a shallow trap (Fig. ~\ref{state4}). The experimental results on photoconductivity of CaF$_2$-Gd crystals show that the ground state of the PC$^+$ center is located at 2 eV below the conductivity band, while PC center produces a deep level at 3.1 eV below the conduction band (Fig. ~\ref{state4}) \cite{Heyman:1969}. Finally, the PC$^+$ - center can be represented as the divalent rare earth ion perturbed by the anion vacancy, and the PC - center is F- center (an electron in the field of anion vacancy) perturbed by the trivalent rare earth ion. Such models are better able to explain the differences in the temperature stability of the PC$^+$ centers. It looks strange that the PC$^+$ centers in CaF$_2$-Y and CaF$_2$-Lu are only stable below room temperature, while PC$^+$ centers with other impurities are stable above room temperature, although the stability of the PC$^+$ center in all cases (according to the existing model) was determined by delocalization of electron from electron traps (usually unknown). Frankly, a similar model of the centers has been earlier proposed in number works by German authors \cite{Bernhardt:1971,Bernhardt:1971a,Bernhardt:1971b}. But the results were hardly notices.

Note that in addition to alkaline earth fluorides, there are only a few crystals in which the divalent states of these impurity ions have been studied. Direct and inverse optical transformations of divalent lanthanum (La$^{2+}$) ions into F centers have been studied in BaFCl-La and SrFCl-La crystals by optical means and EPR. Although the explanation of the authors of \cite{Matsarski:2003, Garcia-Lastra:2007} differs from our interpretation, the similarities between  the photo-induced electron transition from the divalent ion to the anion vacancy are worthy of attention.

A similar photo-induced transition of the electron at 80 K from the impurity ion to the anion vacancy, located in the immediate vicinity of the impurity ion, and vice versa was observed early in NaCl-Mg crystals (Mg$^+$F$\longleftrightarrow$Mg$^0$v$_a$) \cite{Egranov:1982}. Mg - related centers containing anion vacancies in the NaCl-Mg crystal  \cite{Egranov:1982,Egranov:1986}, as well as PC$^+$ centers in alkaline earth fluorides (as above) have features in their temperature dependence of the formation as compared to the conventional form of the centers containing anion vacancy (in particular F-aggregate centers \cite{Tijero:1990}, the formation of which is due to the temperature induced migration of anion vacancies).
  
\section{Positions of the d levels}
The energy required  to transfer an electron from the divalent metal ion to the conduction band can be determined by means of a cycle analogous to that employed in the charge transfer model of excitons (Hilsch and Pohl 1930, Born 1932) \cite{Hilsch:1930, Born:1932}: the metal ion is removed from the cation site, ionized and restored. A similar method has been used to assess the ionization thresholds of divalent rare earth ions in the alkaline earth fluorides \cite{Pedrini:1979,Pedrini:1981,Pedrini:1982a,Pedrini:1986,Dorenbos:2003a}. The net gain in energy  $E_{PI}$  is given by 

\begin{equation}
E_{PI}=-(I-E_m-{\Delta}E_m-E_{pol}-E_A),
\label{eq1}
\end{equation}

where $I$  is the the third ionization potential of the free lanthanide atoms, $E_m$ is the negative electrostatic potential energy at the metal-ion site in the pure crystal (mainly related to the Madelung energy), ${\Delta}E_m$ is the correction for the previous term due to the distortion introduced by the impurity,  $E_{pol}$ is the polarization due to removal of  an electron from the divalent metal ion site, $E_A$ is the electron affinity of the crystal. $E_A$ is relatively low and usually not more than 1 eV, but is  not known well \cite{Pedrini:1986}. The evaluation of electron affinities for many compounds was made by Dorenbos \cite{Dorenbos:2013}. $E_{pol}$ = 1.73, 2.01, and 2.02 eV for CaF$_2$,  SrF$_2$ and BaF$_2$ \cite{Starostin:1973,Starostin:1974}. The values of  $I$ \cite{Martin:1978} and  $E_m$ \cite{Poole:1975}  are known.

\begin{figure}
\centering
\includegraphics[width=5.2in, height=3.6in]{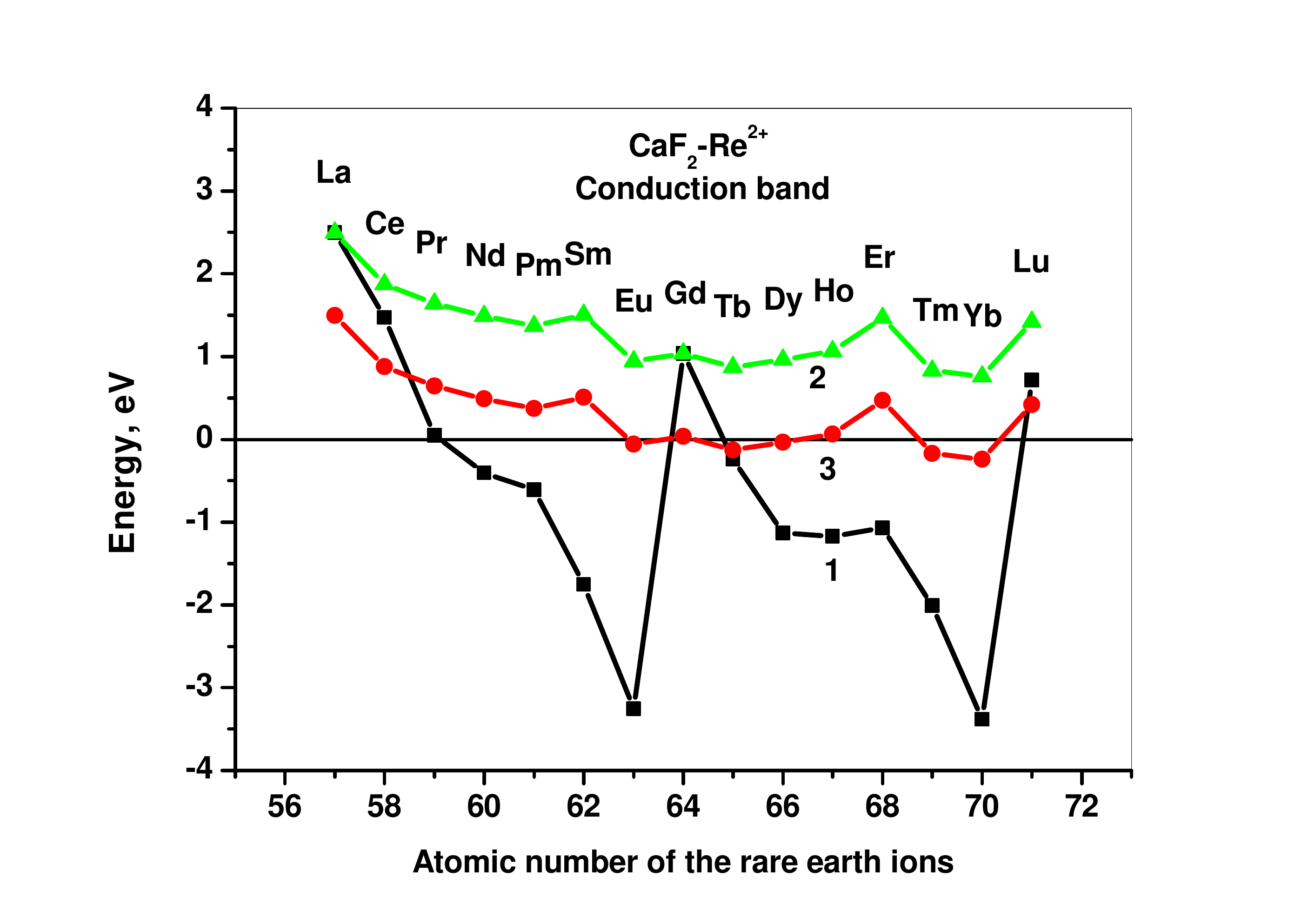}
\caption{The position of the ground state (1), d - level (2) and d(e) -level (3) of divalent ions in relation to the conduction band in CaF$_2$ without ${\Delta}E_m$  correction}
\label{state6}
\end{figure}

In the first stage, it may be preferable  to make evaluation of the ground state of divalent rare earth levels in CaF$_2$ relative to the conduction band without ${\Delta}E_m$  correction due to the distortion introduced by the impurity and without $E_A$. The ground state for La$^{2+}$ and Gd$^{2+}$ is 5d$^1$, for Lu$^{2+}$ is 6s and  f$^n$ is for all others. Figure ~\ref{state6} shows the position of the ground state of divalent ions with respect to the conduction band, which is the same as in the work by Pedrini et al. \cite{Pedrini:1986}. If we use the free ion value for fd (or sd for Lu$^{2+}$) lowest transition energies  \cite{Martin:1978}, the location of d - level respect to the conduction band as shown in figure ~\ref{state6} is obtained. As a rough approximation we take that the lowest energy band d(e$_g$) of the cubic crystal field splitting of the divalent rare earth ion in Ca$^{2+}$ site is located at about 1 eV below initial d level (Fig.~\ref{state6}) \cite{Dorenbos:2002}.

The evaluation shows that the d(e$_g$) level of divalent rare earth ions is located in the conduction band or close to it and the energy position of d(e$_g$) state is significantly less dependent on the type of rare earth ions than the ground state. Chemically the 5d electron possesses similar properties for each lanthanide ion. This means that although the 5d electron has a very strong interaction with the crystal field, the interaction is almost the same for each lanthanide. So sometimes Dorenbos \cite{Dorenbos:2003a,Dorenbos:2004} makes the assumption that the energy difference between the first 5d level and the bottom of the conduction band is constant overall the lanthanides. 

From the experimental data it is known that if the excited d(e$_g$) state of Eu$^{2+}$ or Yb$^{2+}$ is in the conduction band, then it was observed luminescence of these ions with strong Stokes shift which  is called as "anomalous" and that is different in character from the normal d${\rightarrow}$f emission \cite{McClure:1985, Dorenbos:2003a}. It is now accepted that the location of the 5d levels relative to conduction band states and the presence of "anomalous" emission are related to each other. Although at first the anomaly of this emission was explained as arising from the Jahn-Teller distortion \cite{Kaplyanskii:1976}. In CaF$_2$ and SrF$_2$ crystals d(e$_g$) level of Yb$^{2+}$  is localized in the conduction band and an "anomalous" luminescence of these ions has been observed. At the same time d(e$_g$) level of Eu$^{2+}$ is located below the bottom of the conduction band in these crystals and an ordinary luminescence have been observed, but the "anomalous" emission of Eu$^{2+}$  has been observed in BaF$_2$ crystals that indicates the localization of d(e$_g$) level in the conduction band \cite{Pedrini:2007}. Experimental data on the photoconductivity show the position of the ground state of the divalent rare earth ions with respect to the conduction band in alkaline earth fluorides (Table \ref{table}).

\begin{table}
\caption{\label{tab:table1}Positions of the levels of divalent ions in relation to the conduction band from photoconductivity measurements (eV)}
\begin{tabular}{cccccccc}
Re$^{2+}$ & CaF$_2$  & SrF$_2$ & BaF$_2$ &  Re$^{2+}$& CaF$_2$ & SrF$_2$ & BaF$_2$\\
 \hline
La$^{2+}$ &  & & & Gd$^{2+}$ &2.0 \cite{Heyman:1969} & & \\
Ce$^{2+}$ & 1.6 \cite{Pedrini:1981} & 1.3 \cite{Pedrini:1981} & 1.1 \cite{Pedrini:1981} & Tb$^{2+}$ & & &\\
- & 1.9 \cite{Radzhabov:2001} & 1.6 \cite{Radzhabov:2001} & 1.4 \cite{Radzhabov:2001} & - & & &\\
Pr$^{2+}$ &1.3 \cite{Pedrini:1982} & & & Dy$^{2+}$ &1.76 \cite{Pedrini:1979} &&\\
Nd$^{2+}$& & & &Ho$^{2+}$&1.75 \cite{Pedrini:1979} && \\
Pm$^{2+}$ & & & &  Er$^{2+}$&&&\\
Sm$^{2+}$&1.7 \cite{Pedrini:1986} & & & Tm$^{2+}$& 2.75 \cite{Pedrini:1979} & 2.08 \cite{Pedrini:1979} & 1.5 \cite{Pedrini:1979} \\
Eu$^{2+}$ & 3.8 \cite{Pedrini:1986} & 3.8 \cite{Moine:1991} & 2.9 \cite{Moine:1991} & Yb$^{2+}$ & 4.22 \cite{Pedrini:1986} & 3.0 \cite{McClure:1985a} & \\
&&&& Lu$^{2+}$ &&&\\
\end{tabular}
 \label{table}
\end{table}

However, the evaluation shows that the levels of Eu$^{2+}$ and Yb$^{2+}$ are below the bottom of the conduction band, and the level of Yb$^{2+}$ is deeper than that of Eu$^{2+}$ which contradicts the experimental results (Fig.~\ref{state6}).

As can be seen from Figure ~\ref{state6} all the rare earth ions having the ground state 5d$^1$ and forming photochromic centers in alkaline earth fluorides crystals are unstable in the divalent state as their ground state is localized in the conduction band, or at least is close to it (in the case of terbium). 
As mentioned above the divalent rare earth ions (La, Ce, Gd, Tb, and Lu) in various compounds are unstable against electron autodetachment. The formation of the photochromic centers in alkaline earth fluorides (instead of the pure divalent ions) indicates that a similar instability of these divalent ions occurs in the crystals and hence d$^1$ ground state interacts with the conduction band. The behaviour of these divalent ions in alkaline earth fluorides in many respects is similar to that of non-classical divalent rare earth iodides, for which an electron is transferred to the conduction band \cite{Cotton:2006}. 

Some estimations in early studies of Ce$^{2+}$ in CaF$_2$ \cite{Pedrini:1979} and BaF$_2$ \cite{Dorenbos:2003a} gave similar results indicating the chemical instability of the divalent state of the ion. Moreover, Dorenbos \cite{Dorenbos:2003a} pointed out the possibility of stabilization of the Ce$^{2+}$ state due to the nearest defect, which would reduce the level  respect to the conduction band.

\begin{figure}
\centering
\includegraphics[width=7in, height=4.5in]{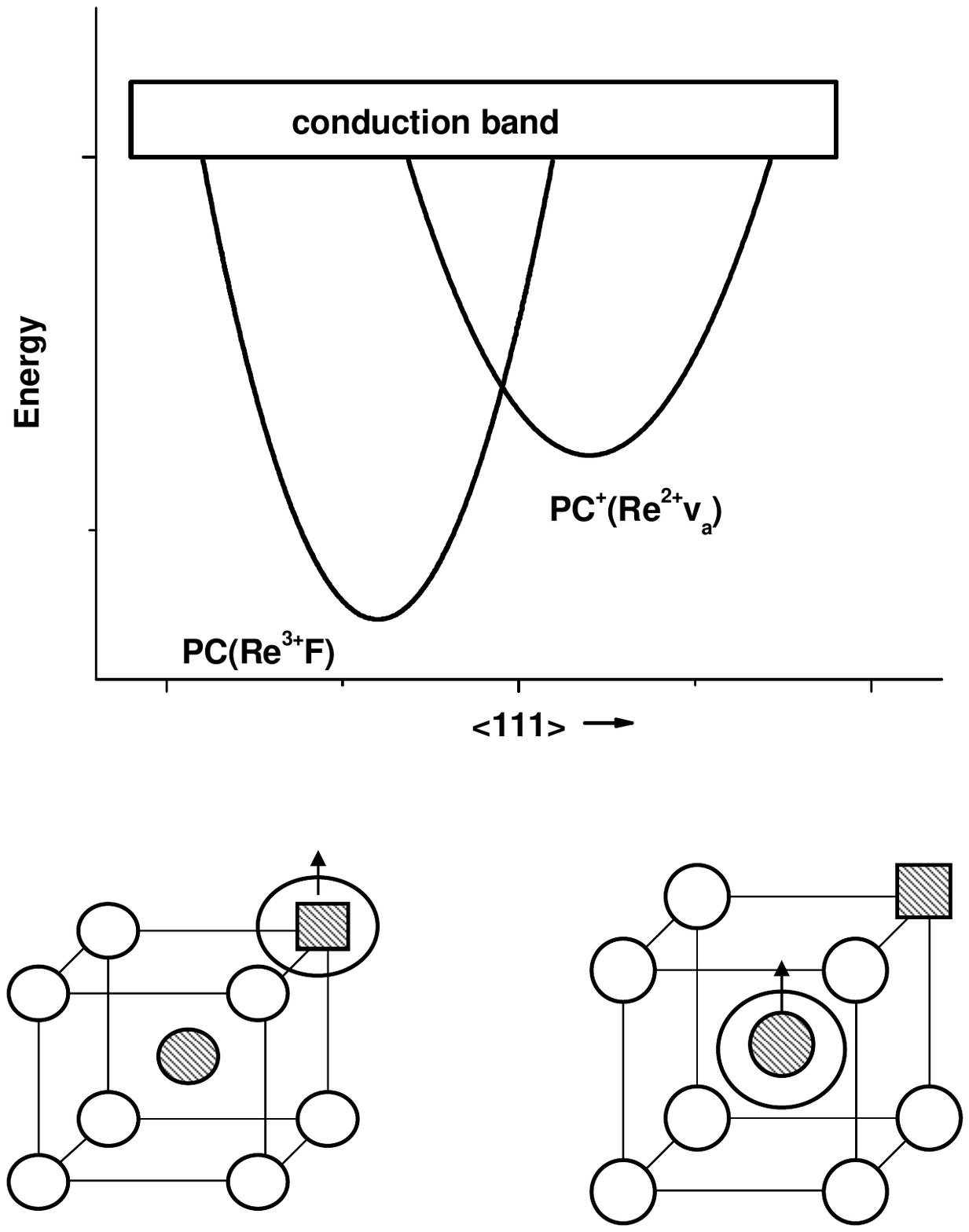}
\caption{Models of PC and PC$^+$ centers and energy scheme of their position in relation to the conductivity zone.}
\label{state5}
\end{figure}

${\Delta}E_m$  depends on  the deformation of  the  lattice  in  the vicinity of the  impurity  and  is  an  important but  not well-known quantity. This is really important, since even ${\Delta}R$=0.01 \AA makes a 0.14 eV change \cite{Pedrini:1979}. It has been shown that the  actual nearest neighbor F$^-$ ion positions around a  divalent or trivalent rare earth ion  can be determined from  the  superhyperfine constants  to  a  good  approximation. The shift in position of the nearest neighbors has an appreciable effect on the Madelung energy at the metal site.  The corrections are calculated simply for the  change  in  the near neighbor distance as was done by Pedrini et al. \cite{Pedrini:1979}. 

\begin{figure}
\centering
\includegraphics[width=5.2in, height=3.6in]{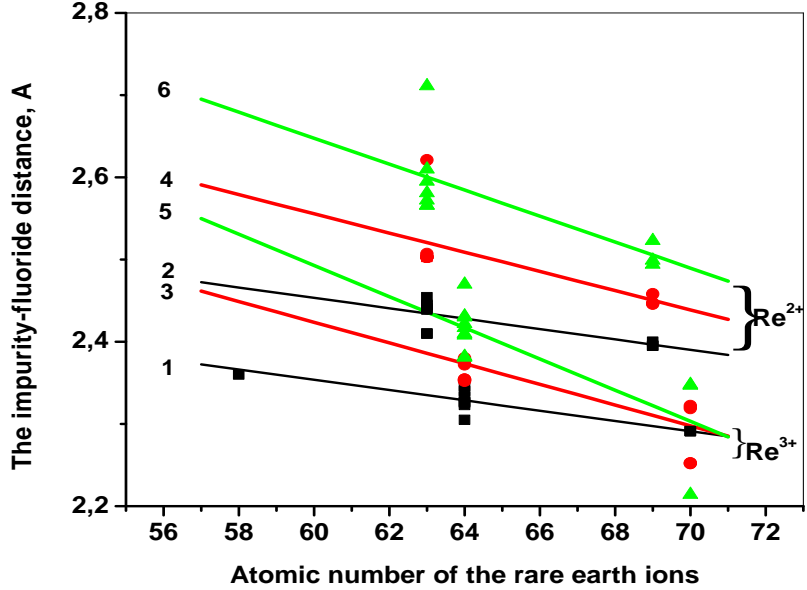}
\caption{Interpolation literary results \cite{Tovar:1983,Baker:1979,Anderson:1975,Baker:1977,Edgar:1975,Baberschke:1971,Baberschke:1972,Fainstein:1982,Gavasheli:2006} are used to estimate the impurity-fluoride distance for the divalent (2,4,6) and trivalent (1,3,5) rare earth ions in CaF$_2$ (1,2), SrF$_2$ (3,4) and BaF$_2$ (5,6) crystals}
\label{state6a}
\end{figure}

The main results on the distortion of the lattice near the impurity  were obtained for divalent ions Eu$^{2+}$ and  Tm$^{2+}$ and for cubic trivalent rare earth ions Gd$^{3+}$ and Yb$^{3+}$ in alkaline earth fluorides \cite{Tovar:1983,Baker:1979,Anderson:1975,Baker:1977,Edgar:1975,Baberschke:1971,Baberschke:1972,Fainstein:1982,Gavasheli:2006}. No distortion was observed for cubic Ce$^{3+}$ in CaF$_2$ \cite{Baker:1968} (Fig. ~\ref{state6a}). The smallest corrections ${\Delta}E_m$ occur for Re$^{3+}$ in the case of CaF$_2$ since the ionic radius of Ca$^{2+}$ is nearest to that of the trivalent rare earths. Many estimates of the lattice distortion by the divalent rare earth ions were made using conventional dependence of the ionic radius in the lanthanide series \cite{Pedrini:1986,Dorenbos:2003a}. The radius is known to decrease moving across the series as a result of decreased shielding of the 4f orbitals in accordance with the lanthanide contraction. However, the ionic radii are known only for traditional divalent rare earth ions with ground state f$^n$ (Nd$^{2+}$, Sm$^{2+}$, Eu$^{2+}$, Dy$^{2+}$, Tm$^{2+}$ and Yb$^{2+}$) \cite{Shannon:1976}. For the rest of the divalent rare earth ions the ionic radii were found by interpolation. However, the remaining rare earth ions (La, Ce, Pr, Gd, Tb, Ho, Er, Lu) have the  5d$^1$ ground state in many compounds \cite{Fieser:2015} and anomalous behavior of the ionic radii in the lanthanide series one would expect.

It recently has been shown that Ln$^{2+}$ ions are accessible for all of the lanthanides except Pm, which was not investigated because of its radioactivity. Complexes of nine new +2 ions (non-classical), La$^{2+}$, Ce$^{2+}$, Pr$^{2+}$, Gd$^{2+}$, Tb$^{2+}$, Ho$^{2+}$, Er$^{2+}$, Y$^{2+}$, and Lu$^{2+}$, were synthesized (\cite{Hitchcock:2008,Meyer:2008,Fieser:2015} and references there in). These new complexes had major structural differences compared to complexes of the traditional six +2 ions in that the difference in bond distances between a +2 ion complex and its +3 ion analog was small. Hence, all of the the non-classical Ln$^{2+}$ complexes have distances that are only 0.020-0.032 \AA  ($\sim$1 \%) longer than their Ln$^{3+}$ analogs. This contrasted so sharply with the 0.10-0.20 \AA ($\sim$6 \%) differences generally seen for complexes of Eu$^{2+}$, Yb$^{2+}$, Sm$^{2+}$, Tm$^{2+}$, Dy$^{2+}$, and Nd$^{2+}$ compared to their Ln$^{3+}$ counterparts. The small increases in bond distances could be explained by the d character in the configurations of the the non-classical +2 ions. There are covalent interactions between the metal d orbitals and the ligands \cite{Fieser:2015}.

It should be noted that using Slater's rules \cite{Slater:1930,Sastri:2003} it can be also obtained that the configuration 4f$^{n-1}$5d$^1$ has a smaller ionic radius than that of 4f$^n$, due to the larger shielding by 4f orbitals (Slater shielding constant = 0.85) than that by 5d orbitals (Slater shielding constant = 0.35), although the difference is not so significant as above. The theoretical results of sixfold and eightfold coordination indicate that the bond shrinkage experienced upon 4f$^n$ to 4 f$^{n-1}$5d$^1$ excitations (5f$^n$ to 5f$^{n-1}$6d$^1$ in the actinides) seems to be a general result of f -element complexes \cite{Barandiaran:2006,Barandiaran:2005}.

\begin{figure}
\centering
\includegraphics[width=5.2in, height=3.6in]{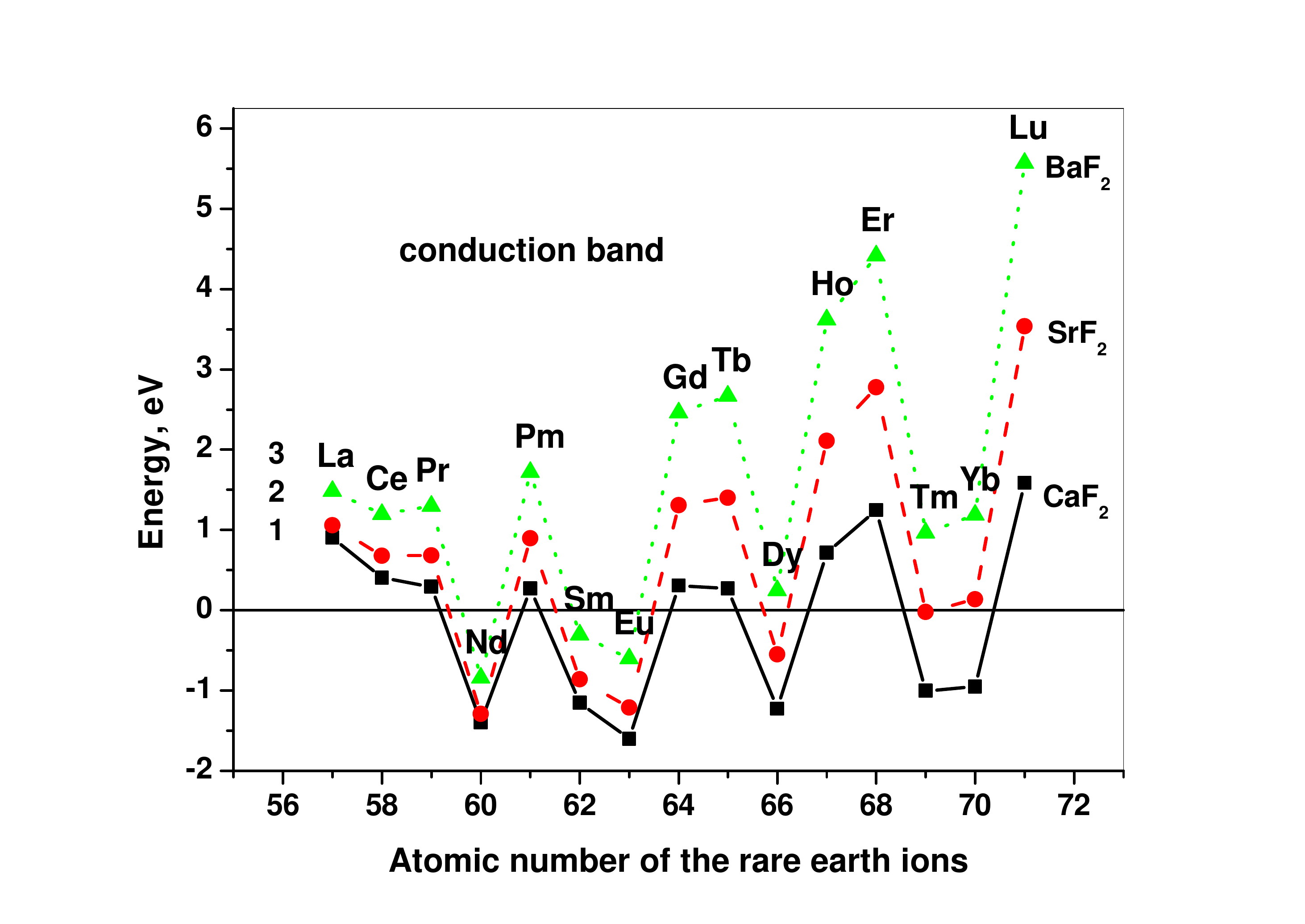}
\caption{The position of the d(e$_g$) level of divalent ions with respect to the conduction band in alkaline earth fluorides: (1) {CaF$_2$}, (2) {SrF$_2$}, (3) {BaF$_2$} with ${\Delta}E_m$  correction}
\label{state6b}
\end{figure}

Given the above results, the lattice distortion around the traditional divalent ions is estimated by interpolating data from Eu$^{2+}$ and Tm$^{2+}$. While for the non-classical divalent ions distortion is estimated by interpolation data from cubic Ce$^{3+}$, Gd$^{3+}$ and Yb$^{3+}$ in CaF$_2$ and shift them upward by 0.02 \AA ( Figure ~\ref{state6a}). It should be noted that the slopes of the curves for the divalent and trivalent ions of the same crystal are similar to each other and the slope in BaF$_2$ is close to the ordinary dependence of the ionic radius of rare earth element (Re$^{3+}$). The final result is shown in Figure ~\ref{state6b}. One can see that d(e$_g$) state for all non-classical divalent ions lies in the conduction band, and for them it is the ground state, it indicates their instability in CaF$_2$ crystals. Due to the large ionic radius of the traditional divalent ions,  d(e$_g$) state of these ions lies below the conduction band, wherein Eu$^{2+}$ is the deepest level, while Yb$^{2+}$ is the shallowest level. If it is assumed that the electron affinity for CaF$_2$ crystals is of about 1 eV, which is not improbable, then d(e$_g$) state of  Yb$^{2+}$ is located in the conduction band, and d(e$_g$) state of Eu$^{2+}$ is below the conduction band bottom, which is consistent with experiments. The similar way is used to make estimates the lattice distortion near the impurity for crystal SrF$_2$ and BaF$_2$ (Figure ~\ref{state6b}).

An evaluation of the ground level of divalent ion relative to the conduction band in the case of the presence in the nearest-neighboring of an anion vacancy gives from the formula ~\ref{eq1} expression, in which the Madelung energy is lowed by the lack of the neighboring fluoride ion:

\begin{equation}
E_{PI}=-(I-E_m-{\Delta}E_m+{e^2\over {\varepsilon}r_0}-E_{pol}-E_A),
\label{eq4}
\end{equation}

where ${\varepsilon}$ is the permittivity (dielectric constant) of the material and $r_0$ is metal-halide distance. In this case, d$^1$ level is lowered relative to the above situation by the $e^2\over {\varepsilon}r_0$. The charged anion vacancy polarizes the lattice around them. There are two extreme cases: the polarization takes place without the displacement of the surrounding ions (high-frequency dielectric constant (${\varepsilon}_\infty$) ) or with their displacement (static dielectric constant (${\varepsilon}_S$)). The table \ref{table2} given values of dielectric constants and the shift of d(e$_g$) level for both cases. A comparison of these results with experimental data on the position of the ground state of divalent ions (table \ref{table}) shows that the polarization occurs without the significant displacement of the ligands \cite{Mott:1938}.

Table ~\ref{table} shows the localization of the ground state of the divalent rare earth ions in relation to the conduction band bottom obtained by the photoconductivity. As it can be see from the table ~\ref{table}, the divalent ions Eu$^{2+}$ and Yb$^{2+}$ have the deepest levels. From Eu$^{2+}$ and Yb$^{2+}$ to La$^{2+}$ and Gd$^{2+}$, the ground state should gradually approach to the conduction band \cite{Dorenbos:2003, Dorenbos:2003a}. However, it is evident that the level for Gd$^{2+}$ is deeper than those for Dy$^{2+}$ and Ho$^{2+}$, and the depth for Ce$^{2+}$ is comparable with that for Sm$^{2+}$ and is is deeper than that of Pr$^{2+}$. 

Thus, the formation of the stable divalent ions as La, Ce, Gd, Tb, Lu, and Y (PC$^+$ centers) in CaF$_2$ and SrF$_2$ crystals during x-ray irradiation occurs via the formation of charged anion vacancies near divalent ions (Re$^{2+}$v$_a$), which lower the d$^1$ ground state of the divalent ion relative to the conductivity band. 

According to the mechanism proposed in our previous paper\cite{Egranov:2013}, there are two necessary conditions for the formation of PC$^+$ centers (Re$^{2+}$v$_a$) by x-ray radiation at low temperature: the ground state is d$^1$ and the localization of its level in the conduction band. It was suggested that instability leading to the formation of anionic vacancies occurs by the localization of the excited d state in the conduction band. This erroneous statement was based on experimental data on the position of the ground state of the divalent ions with respect to the conduction band, in particular, data on Ce$^{2+}$ and Gd$^{2+}$ ions (table \ref{table}). The above results indicate that these experimental results are consistent with the position of the d$^1$ ground state with respect to the conduction band of the divalent ions Ce$^{2+}$ and Gd$^{2+}$ with adjacent anion vacancy. Therefore, the configuration instability leading to the formation of an anion vacancy is not associated with the d$^1$ excited state of the divalent ion, but with the d$^1$ ground state (Figure \ref{state5}).

It should be noted that the processes of the formation of the photochromic centers and the "anomalous" luminescence are similar each other. In both cases the d(e$_g$) orbital in eightfold cubic coordination is split by interaction with E$_g$ distortions of the ligands, and the energy of the entire complex can be lowered by any linear combination of the degenerate E$_g$ mode distortions, Q$_\theta$, and Q$_\varepsilon$ (Jahn-Teller distortion). In both cases the d(e$_g$) level is located in the conduction band. (d(e$_g$) level is the ground state for divalent  La, Ce, Gd, Tb, Lu, and Y ions and is the excited state for Eu$^{2+}$ and Yb$^{2+}$) (Figure \ref{state5}). In our previous article \cite{Egranov:2013}, we thought that quantum interference phenomenon between the discrete d(e$_g$) level and a continuum of conduction band (Fano effect) as well as valence fluctuation creates the resonant electron-phonon interaction that leads to the enhancement the Jahn-Teller distortion. Perhaps the differences between the formation the photochromic centers and the "anomalous" luminescence are in the time interval of both processes. In the case of the "anomalous" emission the process is limited by lifetimes of the luminescence (typically 10$^{-9}$ sec for allowed emission transitions). It is possible that the formation of an anion vacancy can take longer. 

\begin{table}
\caption{\label{tab:table2} High-frequency (${\varepsilon}_\infty$) and static (${\varepsilon}_S$) dielectric constant \cite{Tomiki:1969,Kaiser:1962} and the energy shift (${\Delta}E_a$) of the d(e) level by the anion vacancy}
\begin{tabular}{c|cc|cc}
crystals&${\varepsilon}_\infty$&${\Delta}E_a$ (eV)&${\varepsilon}_S$&${\Delta}E_a$ (eV)\\
 \hline
CaF$_2$ & 2.045& 2.98 & 6.7&0.91\\
SrF$_2$ & 2.07 &2.78 &6.6 & 0.87 \\
BaF$_2$ & 2.6& 2.07&7.33 &0.73 \\
\end{tabular}
 \label{table2}
\end{table}

\begin{figure}
\centering
\includegraphics[width=5.2in, height=3.6in]{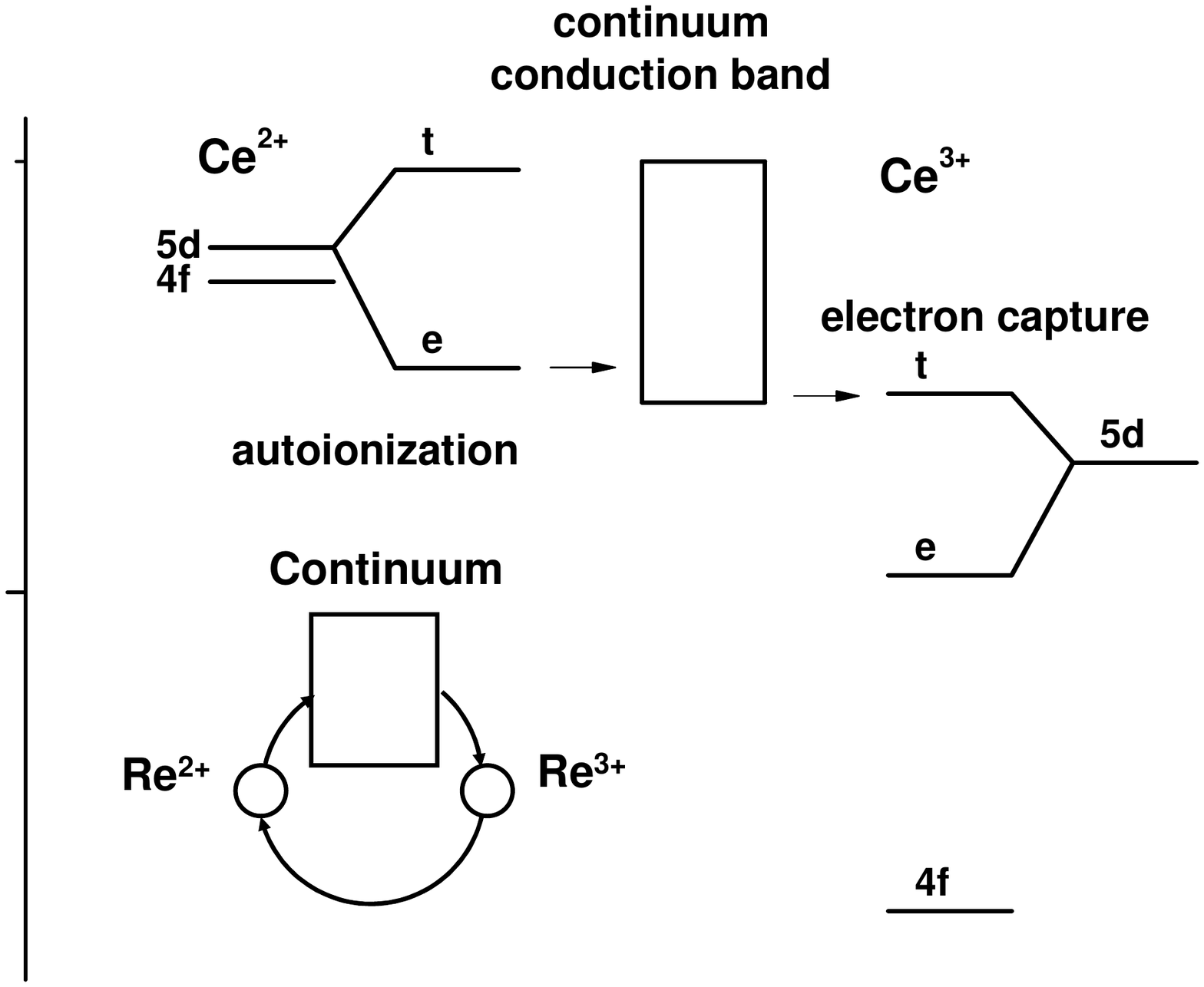}
\caption{Energy levels of Ce$^2$$^+$ and Ce$^3$$^+$ in cubic F$^-$ coordination (schematic) relative to the conduction band in CaF$_2$. Spin-orbit coupling is neglected. The inset shows the cycle associated with the valence fluctuation of the rare earth ion.}
\label{state5}
\end{figure}

\section{Discussion and Conclusion}
As was mentioned above, the non-classical divalent lanthanide complexes (La$^{2+}$, Ce$^{2+}$, Pr$^{2+}$, Gd$^{2+}$, Tb$^{2+}$, Ho$^{2+}$, Er$^{2+}$, Y$^{2+}$, and Lu$^{2+}$) are chemically unstable in many compounds with respect to autodetachment of the electron \cite{Meyer:2012a,Cotton:2006,Meyer:1992}. The divalent rare earth ions (La, Ce, Gd, Tb, Lu, and Y) in alkaline earth fluorides are also unstable with respect to electron autodetachment since its d$^1$(e$_g$) ground state is located in the conduction band \cite{Hulliger:1968}. Localization of doubly degenerate d$^1$(e$_g$) level in the conduction band creates a configuration instability around the divalent rare earth ion that leading to the formation of anion vacancy in the nearest neighborhood \cite{Egranov:2013}.  Thus the formation of the stable divalent ions as La, Ce, Gd, Tb, Lu, and Y (PC$^+$ centers) in CaF$_2$ and SrF$_2$ crystals during x-ray irradiation occurs via the formation of charged anion vacancy Re$^{2+}$v$_a$ near the divalent ion, which lower the ground state of this ion relative to the conductivity band. Photochromic effect occurs under thermally or optically stimulated electron transition from the divalent rare earth ion to the neighboring anionic vacancy. Optically induced reverse electron transition is also possible.

Ligands can act on the impurity divalent rare earth ion in two different directions. On the one hand the ligand set can change the ground electronic state in Ln$^{2+}$ complexes. Apparently, with complexes of the Ln$^{2+}$ ions, the proper ligand  field can lower the energy of the 5d orbitals with respect to the 4f orbitals such that 5d can be part of the ground state. On the other hand the stability of the divalent ion having d$^1$(e$_g$) state roughly determined by the expression (the simplified expression \ref{eq1}):

\begin{equation}
E_{PI}=-(I-E_{ligand}),
\label{eg4}
\end{equation}

where $E_{ligand}$ is the energy of interaction of the impurity ion with ligands. In our case, the Madelung energy $E_m$,  used in the expression \ref{eq1}, can be roughly defined as the energy of the interaction of the impurity ion with ligands of the first two spheres of the nearest environment $E_{ligand}$$\approx$$E_m$. The formation of the anion vacancy reduces the energy of the interaction of the impurity ion  with ligands $E_{ligand}$ and as a result, the formation of the stable divalent rare earth ion. Therefore, the decreasing of the energy $E_{ligand}$ in relation to the third ionization potential $I$ increases the stability of non-classical divalent rare earth ions. The total energy of the metal-ligand interaction $E_{ligand}$ depends on the number of the ligands as well as on the types of interactions (such as ionic or molecular character of the bonding). Perhaps this explains the fact that rare earth ions are La, Ce, Gd, Tb, Lu compounds do not generate a stable divalent state with oxygen, fluorine, chlorine and bromine ions and only for iodides, having the lowest energy $E_{ligand}$ in this series, the compounds with 5d$^1$ electrons delocalized in the conduction band can be produced. Following in this direction, one can assume that the non-classical divalent rare earth compounds with strong ionic bonds would be less stable than the compounds with weaker molecular bonds  that have recently been demonstrated  \cite{Hitchcock:2008,Meyer:2008}.

In analyzing the stability of divalent rare earth ions in the alkaline earth fluorides we used often used division of these ions in the two groups. The first group includes the traditional six rare earth ion (Eu$^{2+}$,Yb$^{2+}$, Sm$^{2+}$, Tm$^{2+}$, Dy$^{2+}$, and Nd$^{2+}$) that readily form the divalent ion complexes. The second  group are the rare earth ions (La$^{2+}$, Ce$^{2+}$, Pr$^{2+}$, Gd$^{2+}$, Tb$^{2+}$, Y$^{2+}$, Ho$^{2+}$, Er$^{2+}$, and Lu$^{2+}$) for which  divalent valence is unusual. However, apparently divalent ions holmium Ho$^{2+}$ are efficiently formed in alkaline earth fluorides by ionizing radiation although the ion is not relate to conventional rare earth ions that forming divalent compounds \cite{Sabisky:1965,Weakliem:1967}. Apparently the ground state of Pr, Ho and Er divalent ions in the alkaline fluorides is 4f$^n$ and they form stable divalent ions in a cubic environment. On the other hand Dy and Nd can form both 4f$^{n+1}$ and 4f$^n$5d$^1$ divalent ions \cite{Fieser:2015} in some complexes.

In conclusion, it should be noted that the similar instability due to the localization of the degenerate d state in the conduction band can be expected for the transition-metal ions.

\section*{Acknowledgments}
This work is supported by RFBR according to the research project No. 15-02-06666a. 

\section*{References}
\bibliographystyle{utphys}
\bibliography{Egranov}

\end{document}